\documentclass{ws-mpla}
\usepackage{graphics}

\def\beq{\begin{equation}}
\def\eeq{\end{equation}}
\def\be{\begin{eqnarray}}
\def\ee{\end{eqnarray}}
\def\ci{\cite}
\def\bi{\bibitem}
\def\lsim{\mathrel{\rlap{\lower4pt\hbox{\hskip1pt$\sim$}}
    \raise1pt\hbox{$<$}}}         
\def\gsim{\mathrel{\rlap{\lower4pt\hbox{\hskip1pt$\sim$}}
    \raise1pt\hbox{$>$}}}         
\def\msun{M_\odot}
\newcommand{\sla}[1]{#1\!\!\!/}

\begin{document}

\markboth{Omar Benhar}
{Neutron star matter equation of state and gravitational wave emission}

%
\catchline{}{}{}{}{}
%

\title{NEUTRON STAR MATTER EQUATION OF STATE\\
AND GRAVITATIONAL WAVE EMISSION}

\author{\footnotesize OMAR BENHAR}

\address{INFN and Department of Physics, Universit\`a ``La Sapienza'',
Piazzale Aldo Moro, 2\\
I-00185 Roma, Italy\\
benhar@roma1.infn.it}

\maketitle

\pub{Received (Day Month Year)}{Revised (Day Month Year)}

\begin{abstract}
The EOS of strongly interacting matter at densities ten to fifteen orders of magnitude 
larger than the typical density of terrestrial macroscopic objects determines a number of 
neutron star properties, including the pattern of gravitational waves emitted
following the excitation of nonradial oscillation modes. This paper reviews some of 
the approaches 
employed to model neutron star matter, as well as the prospects for
obtaining new insights from the experimental study of gravitational
waves emitted by neutron stars.

\keywords{equation of state; neutron stars; gravitational waves.}
\end{abstract}

\ccode{PACS Nos.: 04.30-w, 04.30Db, 97.60Jd.}

\section{Introduction}	
\label{intro}

The equation of state (EOS) is a nontrivial relation linking the
thermodynamic variables that specify the state of a physical system\ci{huang}.
The best known example is Boyle's ideal  gas law, stating
 that the pressure of a collection of $N$ noninteracting, pointlike
classical particles, enclosed in a volume $V$, grows linearly with
the temperature $T$ and the average particle density $n = N/V$.

The ideal gas law provides a good description of very dilute systems, in 
which interaction effects can be neglected. In general, the EOS can be written 
expanding the pressure, $P$,
in powers of the density (throughout this paper, I will use units such that 
$\hbar = c = k_B = 1$, $k_B$ being Boltzmann's constant):
\beq
P = nT\ \left[ 1 + n B(T)  + n^2 C(T) + \ldots \right]\ .
\label{virial}
\eeq
The coefficients appearing in the above series, called
 virial expansion, depend on temperature only, and describe
the departure from the ideal gas law arising from interactions. 
The EOS carries a great deal of dynamical
information and provides a link between measurable {\it macroscopic} quantities, 
such as pressure and temperature, and the forces acting
between the constituents of the system at {\it microscopic} level.

In this brief review, I will discuss theoretical models of strongly interacting matter 
at densities up to fifteen orders of magnitude larger than the typical density of 
terrestrial macroscopic objects. The EOS in this density regime largely determines
neutron star properties, including the pattern of gravitational waves emitted 
following the excitation of nonradial oscillation modes.

Section \ref{ns_structure} provides a summary of neutron star structure, while
some of the approaches employed to model neutron star matter are briefly reviewed in 
Section \ref{EOS}. The dependence of the equilibrium properties of a nonrotating neutron 
star on the EOS describing matter in its interior is discussed in Section \ref{TOV}.
Section \ref{gw} is devoted to the analysis of the imprint of the EOS on gravitational
wave emission. The conclusions are stated in Section \ref{conclusions}. 

\section{Overview of neutron star structure}
\label{ns_structure}

\begin{figure}[ht]
\centerline{\psfig{file=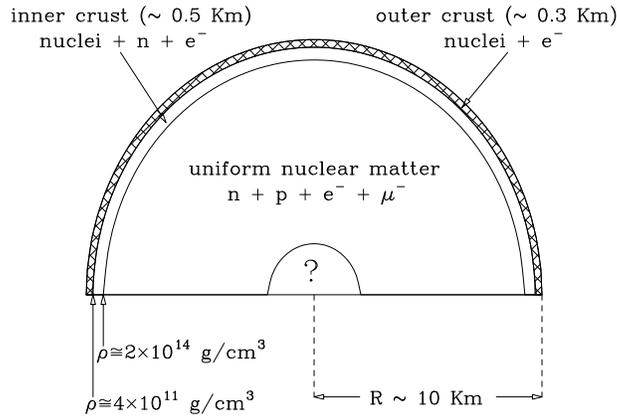,width=3.20in}}
\vspace*{8pt}
\caption{Schematic illustration of a neutron star cross section. Note that 
the equilibrium density of uniform nuclear matter corresponds to 
$\sim 2.7 \times 10^{14} {\rm g/cm}^3$.}
\label{xsec}
\end{figure}

The internal structure of a neutron star, whose cross section is schematically 
illustrated in Fig. \ref{xsec}, is believed to feature a sequence of layers 
of different composition.

The outer crust, about 300 m thick with density ranging from $\rho \sim 10^7 {\rm g/cm}^3$
to the neutron drip density $\rho_d = 4 \times 10^{11} {\rm g/cm}^3$, consists of
a Coulomb lattice of heavy nuclei immersed in a degenerate electron gas.
Moving from the surface toward the interior of the star the density increases,
and so does the electron chemical potential. As a consequence, electron capture becomes
more and more efficient and neutrons are produced in large number through weak 
interactions. 

At $\rho= \rho_d $ there are no more negative energy levels available to the neutrons,
which are therefore forced to leak out of the nuclei.
The inner crust, about 500 m thick and consisting of neutron
rich nuclei immersed in a gas of electrons and neutrons, sets in.
Moving from its outer edge toward the center the density continues to increase,
till nuclei start merging to give rise to structures of variable dimensionality,
changing first from spheres into rods and eventually into slabs.
Finally, at $\rho \sim 2\times 10^{14} {\rm g/cm}^3$, all structures disappear and
neutron star matter reduces to a uniform fluid of
neutrons protons and leptons in weak equilibrium.

The density of the neutron star core ranges between $\sim 2~\times~10^{14}~{\rm g/cm}^3$, at 
the boundary with the inner crust, and a central value that can be as large as
$1 - 4 \times 10^{15}~{\rm g/cm}^3$. Just above $2 \times 10^{14}~{\rm g/cm}^3$ the ground
state of matter
is a uniform fluid of neutrons, protons and electrons. At any given density the
fraction of protons, typically less than $\sim 15 \%$, is determined by the requirements
of weak equilibrium and charge neutrality. At density slightly larger than the equilibrium 
density of uniform nuclear matter, $\rho_0 \sim 2.7 \times 10^{14} {\rm g/cm}^3$, the 
electron chemical potential exceeds the rest mass of the $\mu$ meson and the appearance of 
muons through the the process $n \rightarrow p + \mu^- + {\overline \nu}_\mu$ becomes energetically 
favorable.

All models of EOS based on hadronic degrees of freedom predict that
in the density range $\rho_0 \lsim \rho \lsim 2 \rho_0$ neutron star matter consists 
mainly of neutrons, with the admixture of a small number of protons, electrons and 
muons.

This picture may change significantly at larger density, with the appearance of
 strange baryons produced in weak interaction processes. For example,
although the mass of the $\Sigma^-$ exceeds the neutron mass by more than 250 MeV,
its production becomes energetically allowed as 
soon as the sum of the neutron and electron chemical potentials equals the
$\Sigma^-$ chemical potential.

Finally, as nucleons are known to be composite objects of size $\sim  0.5-1.0$ fm,
corresponding to a density $\sim 10^{15} {\rm g/cm}^3$, it is expected that,
if the density in the neutron star core reaches this value, matter may undergo a 
transition to a new phase, in which quarks are no longer clustered into nucleons 
or hadrons.

\section{Models of neutron star matter EOS}
\label{EOS}

It has long been recognized\ci{OV} that the Fermi gas model, leading to a simple
polytropic EOS, predicts a maximum neutron star mass $\sim 0.7\ \msun$, thus dramatically 
failing to explain the observed neutron star masses\ci{masses}. This failure clearly shows 
that neutron star equilibrium against gravitational colapse requires a 
pressure other than the degeneracy pressure,
whose origin has to be traced back to hadronic interactions.
Unfortunately, the need of including dynamical effects in the EOS is confronted with
the complexity of the fundamental theory of strong interactions, quantum chromo dynamics
(QCD). As a consequence, all available EOS of strongly interacting matter
have been obtained within models, based on the theoretical knowledge of the underlying
dynamics and constrained, as much as possible, by experimental data.

While the properties of matter in the outer crust can be obtained directly from
nuclear data\ci{BPS}, models of the EOS in the inner crust, corresponding to 
$4 \times 10^{11} < \rho < 2 \times 10^{14} {\rm g/cm}^3$, are somewhat based on 
extrapolations of the available empirical information, since the extremely 
neutron rich nuclei
appearing in this density regime are not observed on earth\ci{PRL}. 

However, as most of the neutron star mass resides in the core region 
(typically, the crust olny accounts for less than $\sim$ 2 \% of the total mass),  
the star properties discussed in this review are largely 
unaffected by the details of the EOS describing the crust.
In what follows, I will only discuss the EOS of neutron star matter in the 
region of nuclear and supranuclear density, i.e. at
$\rho \geq \rho_0 = 2.7 \times 10^{14} {\rm g/cm}^3$. 

Note that the results described in the following Sections have been obtained 
under
the two standard assumptions, whose validity has long been established\ci{ST}, that 
i) thermal effects can be disregarded and ii) neutron star matter is transparent to 
neutrinos produced in weak interaction processes.

\subsection{Nucleon matter in $\beta$-equilibrium}

Models of the EOS of neutron star matter at 
$\rho_0 \lsim \rho \lsim 2-3 \rho_0$ rest
on the premise that in this density regime nucleons can still be treated as individual 
particles. This assumption appears to be reasonable, as 
the shape of the proton charge distribution, obtained by Fourier transforming its 
measured elelctric form factor, is such that two protons separated by a distance
of $\sim 1 $ fm hardly overlap one another. 

The EOS of nucleon matter is mainly obtained from two different approaches: 
nonrelativistic nuclear many-body theory (NMBT) and 
relativistic mean field theory (RMFT).

In NMBT, nucleon matter is viewed as a collection of pointlike protons and
neutrons, whose dynamics is described by the nonrelativistic Hamiltonian
\be
H = \sum_i \frac{p_i^2}{2m} + \sum_{j>i} v_{ij} + \sum_{k>j>i} V_{ijk}\ ,
\label{hamiltonian}
\ee
where $m$ and $p_i$ denote the nucleon mass and the momentum of the $i$-th particle, 
respectively, whereas $v_{ij}$ and $V_{ijk}$
describe two- and three-nucleon interactions. The two-nucleon potential, that reduces to
the Yukawa one-pion-exchange potential at large distance, is
obtained from a fit to the available data on the
two-nucleon system, including both deuteron properties and $\sim$ 4000 precisely determined 
nucleon-nucleon scattering phase shifts\ci{WSS}. The purely phenomenological 
three-body term, $V_{ijk} \ll v_{ij}$, is needed to reproduce the binding energies of the
three-nucleon bound states\ci{PPCPW} and the empirical saturation properties of 
symmetric nuclear matter\ci{AP}.

The many-body Schr\"odinger equation associated with the hamiltonian
of Eq.(\ref{hamiltonian}) 
can be solved exactly, using stochastic methods,
for nuclei with mass number $A \leq 10$. The energies
of the ground and low-lying excited states are in excellent agreement with the 
experimental data\ci{WP}. Accurate calculations can also be carried out for uniform
nucleon matter, exploiting translational invariace and using either
a variational approach based
on cluster expansion and chain summation techniques\ci{AP},
or G-matrix perturbation theory\ci{BGLS2000}.

Within RMFT, based on the formalism of quantum field theory, nucleons are
described as Dirac particles interacting through meson exchange\ci{QHD0}.
In the simplest implementation of this approach the
 dynamics, modeled in terms of a scalar and a vector field\ci{QHD1}, is described by 
the lagrangian density
\beq
{\cal L} = {\cal L}_N + {\cal L}_M + {\cal L}_{int}\ ,
\label{def:lag}
\eeq
where 
\beq
{\cal L}_N(x) = {\bar \psi}_N(x) \left( i \sla{\partial} - m \right) \psi_N(x) \ ,
\label{l:N}
\eeq
$\psi_N(x)$ denotes the isospin doublet describing the proton and neutron 
fields and 
\be
\nonumber
{\cal L}_M(x) & = & -\frac{1}{4} F_{\mu\nu}(x)F^{\mu\nu}(x) +
\frac{1}{2} m_\omega^2 V_\mu(x)V^{\mu}(x) \\
&   & \ \ \ \ \ \ \ \ \ \ \ \ \ \ + 
\frac{1}{2} \partial_\mu \phi(x) \partial^\mu \phi(x)
- \frac{1}{2} m_\sigma^2 \phi^2(x) \ ,
\label{l:B}
\ee
with $F_{\mu\nu}(x) = \partial_\nu V_\mu(x) - \partial_\mu V_\nu(x)$.
 The interaction term is defined in such a way 
as to give rise to a Yukawa-like meson exchange
potential in the static limit:
\beq
{\cal L}_{int}(x) = g_\sigma \phi(x) {\bar \psi}_N(x)\psi_N(x) -
g_\omega V_\mu(x) {\bar \psi}_N(x) \gamma^\mu \psi_N(x)\ .
\label{l:int}
\eeq
In the above equations, $V_\mu(x)$ and $\sigma(x)$ denote the vector and scalar 
meson fields, respectively, and $m_\omega$, $m_\sigma$, $g_\omega$ and $g_\sigma$ are 
the corresponding masses and coupling constants.

Unfortunately, the equations of motion obtained minimizing the action turn
out to be tractable only in the mean field approximation, i.e. replacing the 
meson fields
with their vacuum expectation values, which amounts to treating them as classical
fields. Within this scheme the meson masses and coupling constants are determined by 
fitting the empirical properties of nuclear matter, i.e. binding energy, equilibrium 
density and compressibility.

NMBT, while suffering from the obvious limitations inherent in its nonrelativistic nature, 
is strongly constrained by data and has been shown to possess a
highly remarkable predictive power. On the other hand, RMFT is based on a very powerful 
and elegant formalism, but assumes a somewhat oversimplified dynamics, which is not 
constrained by nucleon-nucleon data. In
addition, it is plagued by the uncertainty associated with the use of the mean field
approximation, which is long known to fail in strongly correlated systems\ci{KB}.

Using either NMBT or RMFT one can calculate the energy-density of nucleon matter at 
any baryon number density $n_B$ and proton fraction $x$, as well as the proton 
and nucleon chemical potentials. 

The equilibrium conditions with respect to the processes (neutrinos are omitted, as
they do not contribute to chemical equilibrium)
\beq
n \leftrightarrow p + e^- \ \ , \ \ n \leftrightarrow p + \mu^- \ ,
\eeq
obtained from the minimization of the total energy-density, $\epsilon$, subject 
the constraints of baryon number conservation and charge neutrality, are
\beq
\mu_n = \mu_p + \mu_e \ \ , \ \ \mu_e = \mu_\mu
\label{beta:equil}
\eeq
and
\beq
n_p = n_e + n_\mu \ ,
\label{charge:neutr}
\eeq
where $\mu_i = (\partial \epsilon/\partial n_i)_V$ is the chemical potential of the 
particle of type $i$ ($i = n, p, e, \mu)$, whose number density is denoted by $n_i$. 
Using the definition of the proton fraction, $x=n_p/n_B=n_p/(n_n+n_p)$, 
and exploiting charge neutrality, one can write all chemical potentials 
as a function of $n_B$ and $x$. Thus, for any 
given $n_B$, $x$ is uniquely determined by Eqs.(\ref{beta:equil}).

\begin{figure}[ht]
\begin{center}
\resizebox{0.90\textwidth}{!}{\includegraphics{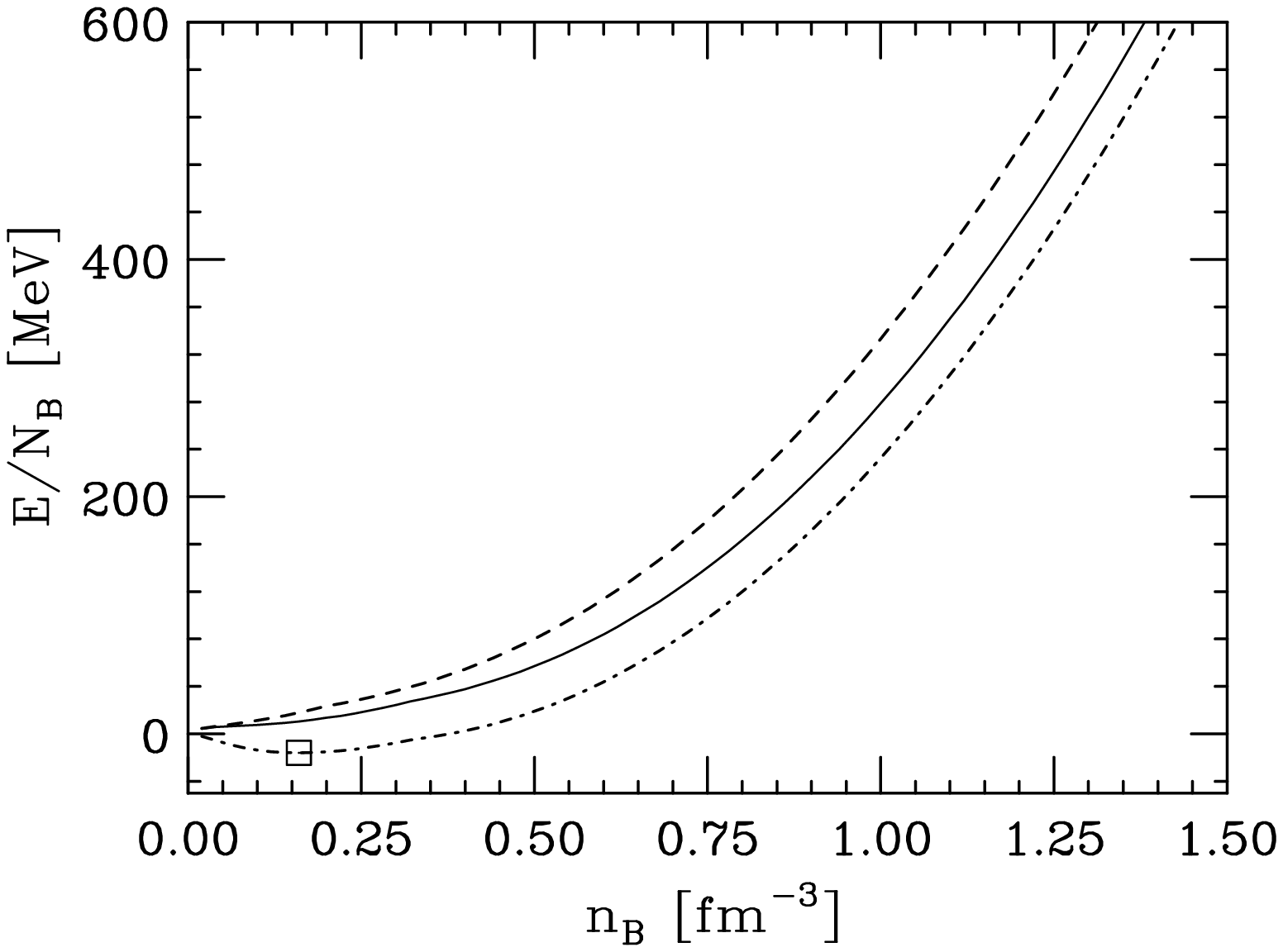}
                              \includegraphics{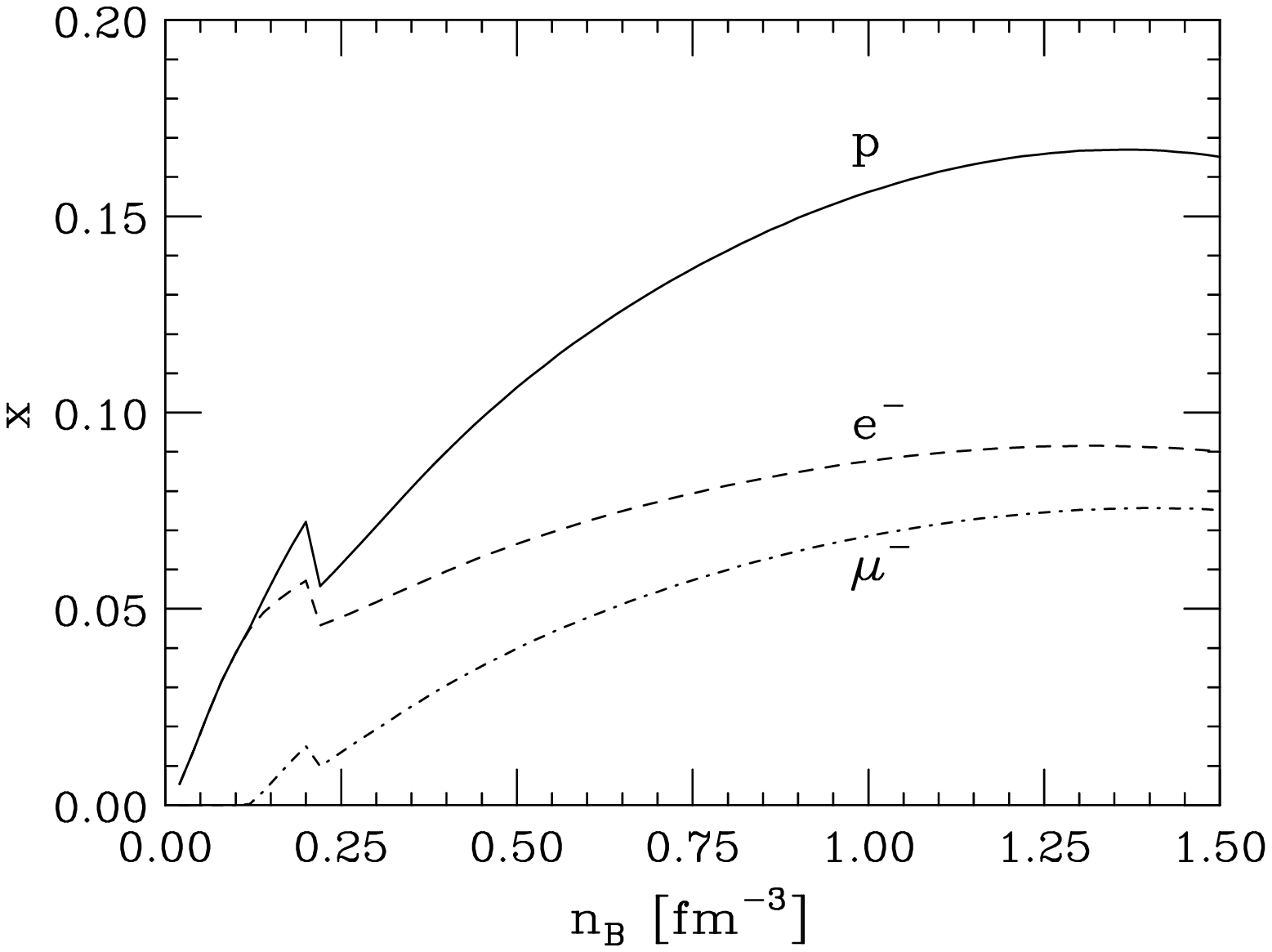}}
\end{center}
\vspace*{8pt}
\caption{Left panel: energy per baryon of nucleon matter calculated by
Akmal et al.\protect\ci{APR}, as a function of baryon number density.
The dashed, dot-dash and solid lines correspond to pure neutron matter,
symmetric nuclear matter and $\beta$-stable matter, respectively.
The box represents the empirical equilibrium properties of symmetric nuclear
matter. Right panel: density dependence of the proton, electron and muon fractions
of $\beta$-stable matter obtained by Akmal et al.\protect\ci{APR}.}
\label{apr}
\end{figure}

Fig. \ref{apr} shows the density dependence of the energy per baryon (left panel) and
the proton, electron and muon  fractions of $\beta$-stable matter calculated by
Akmal et al.\ci{APR} using NMBT. For comparison, the left panel also
includes the results corresponding to pure neutron matter ($x=0$) and symmetric
nuclear matter ($x=1/2$).

The EOS in the form $P = P(\epsilon)$ can be readily obtained from the binding energy
per baryon shown in Fig. \ref{apr} through
\begin{equation}
P = - \frac{1}{n_B^2}\frac{\partial}{\partial n_B}\frac{E}{N_B}\ ,
\label{eos1}
\end{equation}
and
\begin{equation}
\epsilon = n_B \left( \frac{E}{N_B} + m \right) \ .
\label{eos2}
\end{equation}

\subsection{Strange hadronic matter}

As the density increases, different forms of matter, containing
hadrons other than protons and neutrons, may become energetically
favored. For example, the weak interaction process
\beq
n + e^- \rightarrow \Sigma^- + \nu_e\ ,
\label{hypprod}
\eeq
leading to the appearance of a hyperon, sets in as soon as
the condition
\beq
\mu_n + \mu_e  = \mu_{\Sigma^-} 
\eeq
is fulfilled by the chemical potentials (typically at $n_B \gsim 2 n_0$, 
$n_0 = .16\ {\rm fm}^{-3}$ being the baryon number density corresponding to 
 equilibrium of symmetric nuclear matter). At larger density the 
production of $\Lambda$'s is also energetically allowed.

For any given $n_B$, the relative aboundances of the
different hadronic and
leptonic species can be determined from the equations expressing the 
requirements of equilibrium
with respect to weak interactions, conservation of baryon number
and charge neutrality. 

In principle, both NMBT and RMFT can be generalized to take into account the 
appearance of hyperons. However, very little is known of their interactions. 
The available models of the hyperon-nucleon potential\cite{YN} are only loosely
constrained by few data, while no empirical information is available on 
hyperon-hyperon interactions.

\begin{figure}[ht]
\centerline{\psfig{file=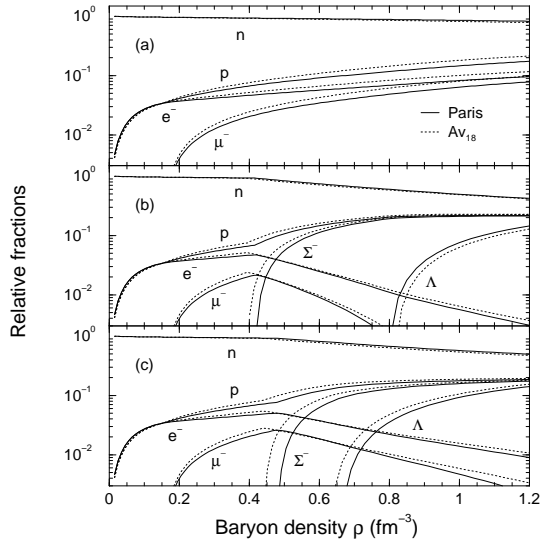,width=2.8in,angle=270}}
\vspace*{8pt}
\caption{Composition of neutron star matter calculated by 
Baldo et al.\protect\cite{hyper} using G-matrix perturbation theory.
Top panel: nucleons and leptons only. Middle panel: nucleons, leptons and
noninteracting hyperons ($\Sigma^-$ and $\Lambda$). Lower panel: hyperon-nucleon
interaction included. The difference between the solid and dashed lines provides 
a measure of the model dependence associated with the NN potential $v_{ij}$
(see Eq. (\protect\ref{hamiltonian})).}
\label{hyperons}
\end{figure}

Fig. \ref{hyperons} shows the composition of neutron star matter resulting from
the NMBT calculations of Baldo et al.\ci{hyper}. Comparison between the middle and
lower panel illustrates the role of hyperon-nucleon interactions, modeled
using the potential of Maessen et al.\cite{YN}, whose inclusion pushes the 
threshold of hyperon production towards larger density. Hyperon-hyperon
interactions are neglected altogether.

The transition from nucleon matter to hadronic matter 
makes the EOS softer, thus leading to a larger compressibility. 
This feature can be easily understoood, as processes like the one
of Eq. (\ref{hypprod}) replace particles carrying large Fermi energies 
with more dilute, and therefore less energetic, strange baryons.

\subsection{Quark matter}

Due to the complexity of QCD, a first principle description of the EOS of 
quark matter at high density and zero temperature is out of reach of the
existing computational approaches. Following the pioneering work of Baym and 
Chin\ci{baym}, a number of authors have carried out numerical studies of quark 
matter based on the simple MIT bag model\ci{bagmodel}. Within this model the 
main features of QCD, namely
confinement and asymptotic freedom, are implemented through the assumptions that:
i) quarks occur in color neutral clusters confined to a finite region of space (the bag),
whose volume is limited by the pressure of the QCD vacuum (the bag constant $B$),
and ii) residual interactions between quarks are weak, and can be treated
in low order perturbation theory.

Neglecting quark masses, the bag model EOS, at first order in the color 
coupling constant $\alpha_s$, can be obtained in closed form 
from the relations linking pressure and energy-density to the quark chemical 
potentials
\beq
P = \frac{1}{4\pi^2} \left( 1 - \frac{2\alpha_s}{\pi} \right)
\sum_f \mu_f^4 - B \ ,
\eeq
and
\beq
\epsilon = \frac{3}{4\pi^2} \left( 1 - \frac{2\alpha_s}{\pi} \right)
\sum_f \mu_f^4 + B \ ,
\eeq
where the sum runs over the active flavors, typically $u$, $d$ and $s$, and $\mu_f$
is the chemical potential of the quark of flavor $f$.

For any baryon desity, quark densities are dictated by the requirements of
baryon number conservation, charge neutrality and weak equilibrium.
In the case of three active flavors one finds
\beq
n_B = \frac{1}{3} (n_u + n_d + n_s) \ \ \ , \ \ \  
\frac{2}{3} n_u - \frac{1}{3}n_d - \frac{1}{3}n_s - n_e = 0
\label{bnc:3}
\eeq
and
\beq
\mu_d  = \mu_s = \mu_u + \mu_e\ .
\label{we:3}
\eeq
Note that in the above equations the possible appearance of $\mu$ mesons is not 
taken into account, as
the results of numerical calculations show that the electron chemical potential 
never exceeds the muon mass.

\begin{figure}[ht]
\centerline{\psfig{file=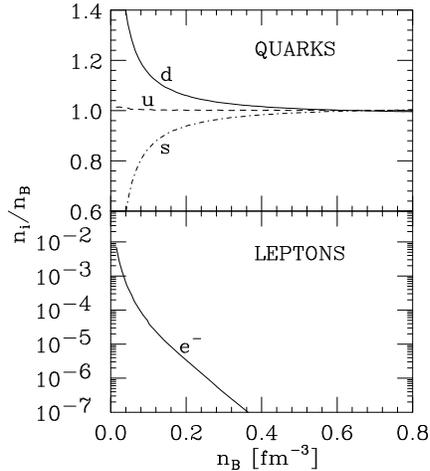,width=2.20in,angle=000}}
\vspace*{8pt}
\caption{Composition of charge neutral matter of $u$, $d$ and $s$ quarks and electrons
in weak equilibrium, obtained from the MIT bag model setting $m_u=m_d=0$, 
$m_s=150$ MeV, $B=200$ MeV/fm$^3$ and $\alpha_s=0.5$.}
\label{quarkmat}
\end{figure}

Fig. \ref{quarkmat} shows the composition of charge neutral quark matter in weak equilibrium. 
 obtained from the MIT bag model. The calculations have been carried out including massless
$u$ and $d$ quarks and strange quarks of mass $m_s=150$ MeV, and
setting $B=200$ MeV/fm$^3$ and $\alpha_s=0.5$. It appears that at large densities quarks of 
the three different flavors are present in equal number, and leptons are no longer needed 
to guarantee charge neutrality.

\section{EOS and properties of nonrotating neutron stars}
\label{TOV}

Plugging the EOS, $P=P(\epsilon)$, into the Tolman Oppenheimer
Volkoff (TOV) equations\cite{T,OV}
\begin{equation}
\frac{dP(r)}{dr} = - G\
\frac{ \left[ \epsilon(r) + P(r) \right]
 \left[ M(r) + 4 \pi r^2 P(r) \right] }
{ r^2 \left[ 1 - 2 G M(r) /r \right] }\ ,
\label{TOV1}
\end{equation}
where G denotes the gravitational constant, and
\begin{equation}
M(r) = 4 \pi \int_0^r {r^\prime}^2 d{r^\prime}
\epsilon({r^\prime})\  ,
\label{TOV2}
\end{equation}
one can obtain the properties of stable nonrotating neutron stars.
Eqs. (\ref{TOV1}) and (\ref{TOV2}) are solved by integrating outwards with the initial
condition $\epsilon(r=0) = \epsilon_c$. For any given value of the central
density, $\epsilon_c$, the star radius $R$ is determined by the condition $P(R)=0$ and
its mass $M=M(R)$ is given by Eq. (\ref{TOV2}).

Plotted as a function of central density, the neutron star mass exhibits a maximum,
 whose value $M_{max}$ is mostly determined by the stiffness of
the EOS. Stiffer EOS, corresponding to more incompressible neutron star matter,  
lead to larger $M_{max}$. Therefore, in principle, measurements of neutron star masses 
may be used to constrain the models of EOS at $\rho>\rho_0$.

Unfortunately, comparison of the calculated $M_{max}$ with the neutron star masses 
obtained from observations, ranging between $\sim$ 1.1 and 
$\sim$ 1.9 $\msun$\cite{masses,masses2}, does not provide a stringent test on the 
EOS, as most
models predict a stable star configuration with mass compatible with the data.
A stronger constraint may soon come from measurements of the neutron star mass-radius ratio.
It has been recently reported\cite{cottam2002} that the Iron and Oxygen transitions observed 
in the spectra of 28 bursts of the X-ray binary EXO0748-676 correspond to a gravitational
redshift $z=0.35$, yielding in turn a mass-radius ratio of the 
source $M/R = 0.153\ \msun/{\rm km}$.

\begin{figure}[ht]
\centerline{\psfig{file=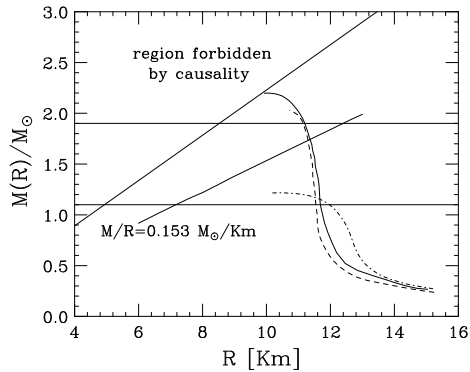,width=2.40in,angle=000}}
\vspace*{8pt}
\caption{
Mass radius relation for different EOS. Solid line: nucleons only\protect\cite{AP}. 
Dotdash: nucleons and hyperons\protect\cite{hyper}. Dashes: nucleons and deconfined 
quarks\protect\cite{BR}. The horizontal lines denote the observational limits on the 
neutron star mass, whereas the other straight lines correspond to the gravitational 
redshift measurement of Cottam et al.\protect\cite{cottam2002} and to the boundary 
of the region forbidden by causality.
}
\label{redshift}
\end{figure}

The results of Cottam et al.\cite{cottam2002} are still
somewhat controversial and need to be confirmed. However, Fig. \ref{redshift} shows that the 
$M(R)$ relations corresponding to EOS obtained from NMBT including only nucleon degrees 
of freedom are consistent with the redshift measurement. The possible appearance of 
deconfined quark matter in a small region in the center of the star does not dramatically 
change the picture\cite{BR}, while the occurrence of a transition to hyperonic matter 
at densities as low as twice the 
equilibrium density of nuclear matter seems to be ruled out\cite{astero}.
From Fig. \ref{redshift} it also appears that none of the considered EOS, all of
them obtained from nonrelativistic NMBT, predicts a $M(R)$ curve extending 
into the region, forbidden by causality, in which the speed of sound in matter exceeds
the speed of light\cite{lattimer}.

\section{Gravitational waves from neutron stars}
\label{gw}

When a neutron star is perturbed by some external or internal event, it can be set
into non radial oscillations, emitting gravitational
waves at the charateristic frequencies of its quasi-normal modes.
This may happen, for example, as a consequence of a glitch,
a close interaction with an orbital companion, a phase transition occurring in 
its inner core or in the aftermath of a gravitational collapse.
The frequencies and damping times of the quasi-normal modes 
 carry information on the structure of the star and the properties of matter 
in its interior. 

Quasi-normal modes are classified according to the source of the restoring 
force which prevails in bringing the perturbed element of fluid back to the 
equilibrium position. Thus, we have a g-mode if the restoring force is mainly 
provided by buoyancy, or a p-mode if it is due to a gradient of pressure.
The frequencies of the g-modes are lower than those of the p-modes,
the two sets being separated by the frequency of the fundamental f-mode, 
associated with global oscillations of the fluid.
General relativity also predicts the existence of additional modes, called w-modes, which 
are purely gravitational, as they do not induce fluid motion\cite{CF1991,KS1992}.
The w-modes can be both polar (even parity) and axial (odd parity). They are highly 
damped and, in general, their frequencies are higher than the p-mode frequencies.

\begin{figure}[ht]
\vspace*{-16pt}
\centerline{\psfig{file=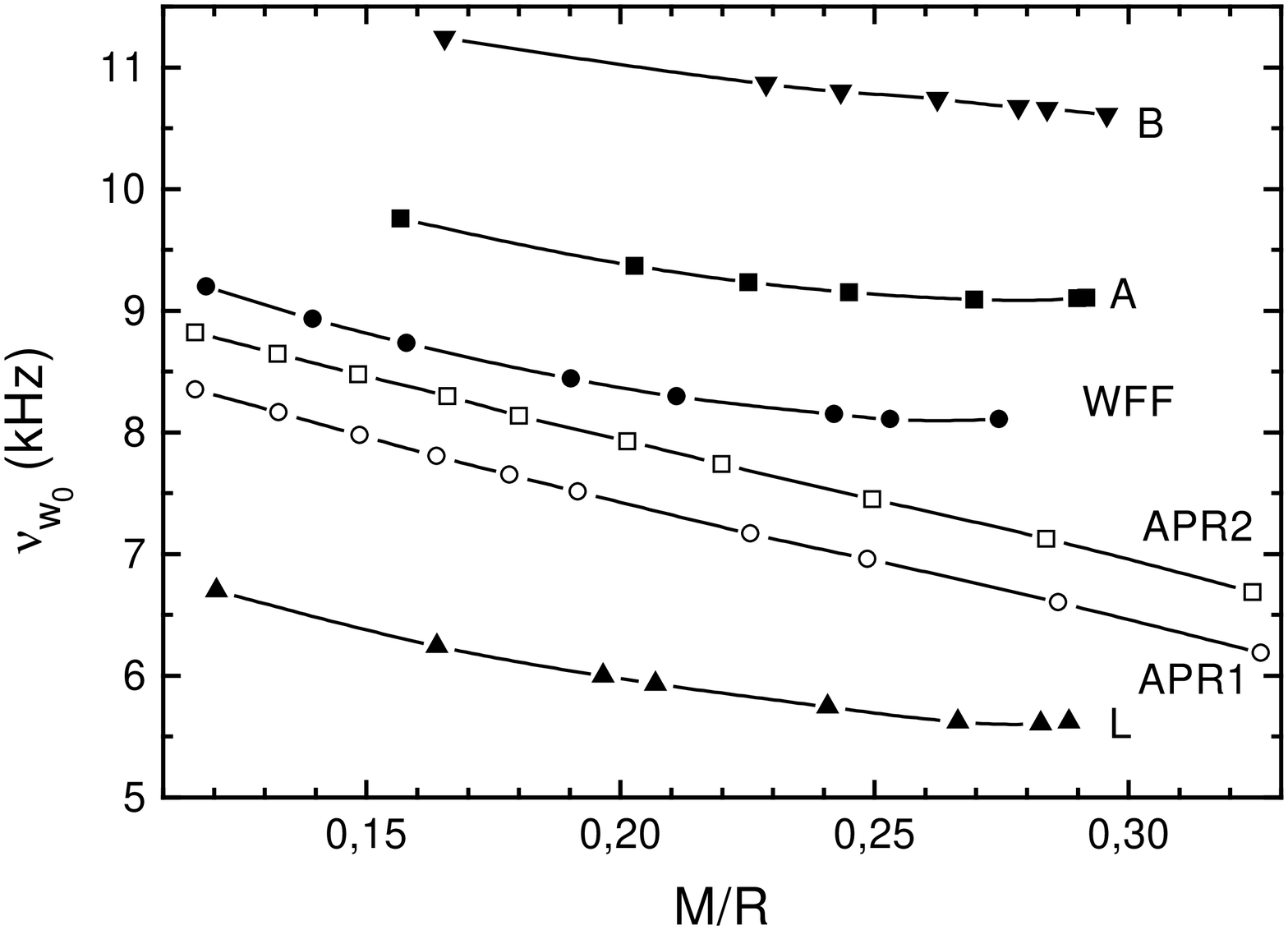,width=3.00in,angle=000}}
\vspace*{-8pt}
\caption{
Frequencies of the first axial w-mode of a nonrotating neutron star, plotted as a 
function of the star compactness. The different curves correspond to different models 
of EOS\protect\cite{mnras}.
}
\label{gw1}
\end{figure}

To see an example of how the oscillation frequencies of quasi-normal modes 
depend on the neutron star matter EOS, consider 
the axial w-modes of a nonrotating star. In this case, the complex frequencies 
are eigenvalues of a Schr\"odinger-like equation whose potential $V_\ell(r)$
explicitely depend upon the EOS according to\cite{Paper2}
\beq
V_\ell(r) = \frac{ {\rm e}^{2 \nu(r)} }{r^3}\ \left\{
\ell(\ell+1)r + r^3 \left[ \epsilon(r) - P(r) \right] - 6 M(r) \right\}\ ,
\eeq
where
\beq
\frac{d\nu}{dr} = - \frac{1}{\left[ \epsilon(r) + P(r) \right]}
\frac{dP}{dr}\ .
\eeq
Using a set of different EOS one obtains strongly damped eigenmodes\cite{mnras}, 
whose frequencies exhibit the behavior displayed in Fig. \ref{gw1}.

The pattern emerging from Fig. \ref{gw1} strictly reflects the stiffness of the different
EOS in the relevant density region (typically $\rho_0< \rho < 5\rho_0$), softer EOS
correponding to higher frequencies. For example, the curve labelled B
has been obtained from a model based on NMBT and including strange baryons\cite{EOSB}, 
which leads to a very soft EOS and a maximum mass of $\sim$ 1.4 $M_\odot$. On the other hand, 
the curve labelled L corresponds to a RMFT calculation including nucleons 
only\cite{EOSL}, 
yielding a maximum mass of $\sim$ 2.7 $M_\odot$.

In 1998, Andersson and Kokkotas\cite{AK} computed the frequencies of the f-mode, 
the first p-mode and the first polar w-mode of a non rotating neutron star for 
a number of EOS available at that time. They fitted the results of their calculations 
with appropriate {\it universal functions} of the macroscopic properties of the
star, i.e. mass and radius, and showed how these empirical relations
could be used to put constraints on these quantities if the frequency
of one or more modes could be identified in a detected gravitational
signal.

\begin{figure}[ht]
\centerline{\psfig{file=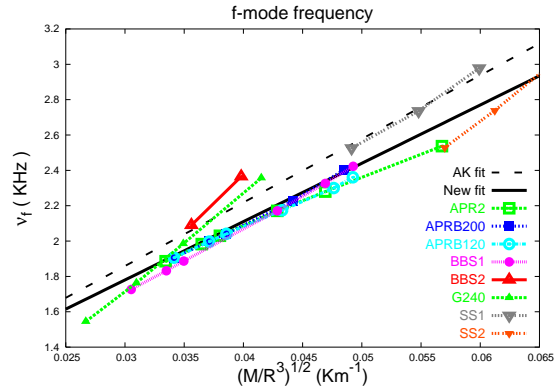,width=2.00in,angle=270}}
\vspace*{8pt}
\caption{Frequency of the f-mode plotted as a function of the square root of the average 
density for different EOS\protect\cite{astero}.
The fits of Andersson and Kokkotas\protect\cite{AK} and Benhar et al.\protect\cite{astero} 
are labelled AK fit and New fit, respectively.}
\label{gw2}
\end{figure}

A similar analysis has been recently carried out by Benhar et al.\ci{astero}, who 
have included in their fit results obtained from state-of-the art EOS. 
As an example, Fig. \ref{gw2} shows the frequencies of the fundamental mode 
corresponding to different EOS, plotted as as a function of the square root of the 
average density. It appears that the frequencies corresponding to different EOS tend 
to {\it scale} to a straight line. The sizable displacement
($\sim$ 100 Hz) of the scaling line of Benhar et al. (thick solid straight line), 
with respect to the one obtained by Andersson and Kokkotas\cite{AK} (dashed straight line),
reflects more than a decade of improvements of theoretical models of the EOS.

The damping time of the f-mode and the frequencies and damping times of the p- 
and w-modes also exhibit scaling, when plotted as a function of the compactness $(M/R)$. 
As a consequence, as pointed out by Anderson and Kokkotas\cite{AK,AK2}, identification 
of the frequency and damping time of some of these modes in a detected 
gravitational signal would allow one to determine the radius of the source knowing 
its mass.

In addition to carrying out the analysis proposed by Anderson and Kokkotas\cite{AK,AK2}, 
one may want to address a different, more direct, question. Will the simultaneous 
knowledge of the mode frequencies and the mass of the star, its only accurately measured 
 observable, provide a severe test to discriminate among different EOS?

As an example, let us consider the fundamental mode. Numerical simulations show that 
this is the mode which is mostly likely to be excited in many astrophysical processes 
and its damping time is quite long, so that it should appear as a sharp
peak in the spectrum of the gravitational signal.

\begin{figure}[ht]
\centerline{\psfig{file=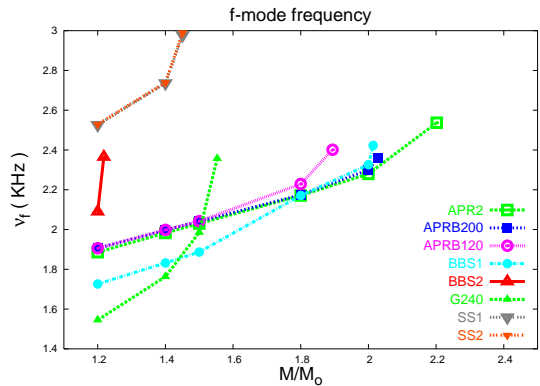,width=2.00in,angle=270}}
\vspace*{8pt}
\caption{Frequency of the f-mode, plotted as a function of the mass of the emitting
star for different models of EOS\protect\cite{astero}. }
\label{gw3}
\end{figure}

Figure \ref{gw3} shows the mass dependence of the f-mode frequencies 
obtained from the EOS models included in the 
analysis of Benhar et al.\cite{astero}.

Comparison between the curve labelled APR2 and those labelled APRB120 and APRB200
indicates that the presence of quark matter in the star inner core does not significantly
affect the pulsation properties of the star. This is a general feature,
also observed in the behavior of p- and w-modes.

The BBS1 and APR2 EOS, based on similar dynamical models, turn out 
to yield appreciably different
f-mode frequencies. This is likely to be ascribed to the effect of the
relativistic corrections included in the APR2 EOS and to
different treatments of three-nucleon interactions.

The transition to hyperonic matter, predicted by the BBS2 model, produces a sizable
softening of the EOS,
thus leading to stable neutron star configurations of very low mass. As a consequence, 
the corresponding f-mode frequency is significantly higher than those obtained from
the other models. So much higher, in fact, that its detection would provide a clear 
signature of the presence of hyperons in the neutron star core.

It is also interesting to compare the f-mode frequencies corresponding to models BBS2 
and G240, as they both predict the occurrence of strange baryons but are obtained from
different theoretical approaches, NMBT and RMFT, based on different 
descriptions of the underlying
dynamics. The behavior of $\nu_f$ displayed in Fig. \ref{gw3} directly reflects the 
relations between
mass and central density obtained from the two EOS, larger frequencies being
always associated with larger densities. For example, the configurations
of mass  $\sim$ 1.2 $\msun$ obtained from the G240 and BSS2 have central
densities $\sim$ 7$\times$10$^{14}$ g/cm$^3$ and $\sim$ 2.5$\times$10$^{15}$ g/cm$^3$,
respectively.  On the other hand, the G240 model requires a central density
of $\sim$ 2.5$\times$10$^{15}$ g/cm$^3$ to reach a mass of $\sim$ 1.55 $\msun$ and 
a consequent $\nu_f$ equal to that of the BBS2 model.

The curves labelled SS1 and SS2 correspond to {\it strange stars}, entirely made of 
quark matter with equal number of up, down and strange quarks.
The peculiar properties of these stars, which are also apparent in the mass-radius 
relation, largely depend on their being self-bound objects\cite{strange1,strange2}.

\section{Conclusions}
\label{conclusions}

The results briefly reviewed in Section \ref{gw} suggest that  
the observation of gravitational waves emitted by neutron stars may
provide valuable new insight, allowing one to severely constrain theoretical
models of strongly interacting matter at high density and zero temperature. 

To address the question of whether the existing gravitational antennas may actually 
be able to detect these signals
let us consider, as an example, a neutron star of mass $M=1.4~ M_\odot$ described 
by the EOS of Akmal et al.\ci{APR}, and assume that
its fundamental mode (whose frequency and damping time are $\nu_f=1983~Hz$ and 
$\tau_f=0.184~s$, respectively\cite{astero}) has been excited by 
some external or internal event.
The signal emitted by the star can be modeled as a damped sinusoid\cite{Kyoto2002}
\be
 h (t) = {\cal{A}} e^{(t_{\rm arr} -t)/\tau_f} \sin
        \left[2\pi \nu_f \left(t - t_{\rm arr}\right)\right]\ ,
\label{signal}
\ee
where $t_{\rm arr}$ is the arrival time and ${\cal A}$ is the mode amplitude.
The energy stored into the  mode can be estimated by integrating the energy flux
\be
\label{ene}
\frac{dE_{\rm{mode}}}{dS d\nu} = \frac{\pi}{2}\,\nu^2\,|\,\tilde h(\nu )\,|^2 ,
\ee
where ${\tilde h}(\nu )$ is the Fourier transform of $h(t)$, over surface ($S$) and 
frequency ($\nu$). 

The signal to noise ratio (SNR) can be expressed in terms of ${\tilde h}(\nu )$
and the noise power spectral density of the detector, $S_n(\nu)$. In the case 
of the ground based interferometric antenna VIRGO\cite{DIS1} one finds 
that SNR=5 corresponds to
$E_{f} \sim 6\times 10^{-7}~M_\odot$ for a source in our Galaxy (distance
from Earth $d\sim 10 ~{\rm kpc}$) and  $E_{f} \sim 1.3 ~M_\odot$ for a source in
the VIRGO cluster ($d\sim 15 ~{\rm Mpc}$). Similar estimates can be found  for the
detectors LIGO, GEO and TAMA.

These numbers suggest that it is unlikely that the first generation of interferometric  
antennas will detect gravitational waves emitted by an oscillating  neutron star.
However, detection will become possible with the next generation of detectors, 
which are expected to be much more sensitive at frequencies above 1-2 kHz.

Identification of the modes in a detected signal, combined with the
 knowledge of the mass of the emitting star will provide 
critical new information on its internal structure, thus opening 
the era of {\it gravitational wave asteroseismology}. 

\section*{Acknowledgments}

This brief review is based on a talk given at the W.K. Kellog Radiation Laboratory, 
California
 Institute of Technology, whose hospitality is gratefully acknowledged. The 
results discussed in Sections \ref{TOV} and \ref{gw} have been obtained in collaboration 
with E. Berti, V. Ferrari, L. Gualtieri and R. Rubino.


\begin{thebibliography}{0}

\bibitem{huang}
K. Huang {\it Statistical Mechanics} (Wiley, New York, 1963).

\bibitem{OV}
J.R. Oppenheimer and G.M. Volkoff, {\it Phys. Rev.} {\bf 55}, (1939) 374.

\bibitem{masses}
S.E. Thorsett and D. Chakrabarty, {\it ApJ} {\bf 512}, 288 (1999).

\bibitem{BPS}
G. Baym, C.J. Pethick and P. Sutherland,  {\it ApJ} {\bf 170}, 299 (1971).

\bibitem{PRL}
C.J. Pethick, B.G. Ravenhall and C.P. Lorenz, {\it Nucl. Phys.} {\bf A584}, 
675 (1995).

\bibitem{ST}
S.L. Shapiro and S.A. Teukolsky, {\it Black Holes, White Dwarfs and
Neutron Stars} (Wiley, New York, 1983).

\bibitem{WSS}
R.B. Wiringa, V.G.J. Stoks and R. Schiavilla, {\it Phys. Rev. C} {\bf 51}, 
38 (1995).

\bibitem{PPCPW} 
B.S. Pudliner, V.R. Pandharipande, J. Carlson, S.C. Pieper
 and R.B. Wiringa, {\it Phys. Rev. C} {\bf 56}, 1720 (1995).

\bibitem{WP}
S.C. Pieper and R.B. Wiringa, {\it Ann. Rev. Nucl. Part. Sci.}
{\bf 51}, 53 (2001).

\bibitem{AP}
A. Akmal and V. R. Pandharipande, {\it Phys. Rev. C} {\bf 56}, 2261 (1997).

\bibitem{BGLS2000}
M. Baldo, G. Giansiracusa, U. Lombardo and H.Q. Song,
{\it Phys. Lett.} {\bf B473}, 1 (2000).

\bibitem{QHD0}
B.D. Serot and J.D. Walecka, {\it Adv. Nucl. Phys.} {\bf 16}, 1 (1986).

\bibitem{QHD1}
J.D. Walecka, {\it Ann. Phys.} {\bf 83}, 491 (1974).

\bibitem{KB} 
L.P. Kadanoff and G. Baym, {\it Quantum Statistical Mechanics}
(Benjamin, New York, 1972).

\bibitem{APR}
A. Akmal, V.R. Pandharipande and D.G. Ravenhall, {\it Phys. Rev. C} {\bf 58}, 
1804 (1998).

\bibitem{hyper}
M. Baldo, G.F. Burgio and H.-J. Schulze, {\it Phys. Rev. C} {\bf 61}, 
055801 (2000).

\bibitem{YN}
P.M.M. Maessen, T.A. Rijken and J.J. deSwart, {\it Phys. Rev. C} {\bf 40}, 
2226 (1989).

\bibitem{baym}
G. Baym and S.A. Chin, {\it Phys. Lett.} {\bf B62}, 241 (1976).

\bibitem{bagmodel}
A. Chodos, R.L. Jaffe, K. Johnson, C.B. Thorne and V.F. Weiskopf, 
{\it Phys. Rev. D} {\bf 9}, 3471 (1974).

\bibitem{T}
R.C. Tolman, {\it Relativity, Thermodynamics and Cosmology}, (Oxford
University Press, 1934)

\bibitem{masses2}
H. Quaintrell, {\it et al.}, {\it A\&A}, {\bf 401}, 303 (2003).

\bibitem{cottam2002}
J. Cottam {\it et al}, {\it Nature}, {\bf 420}, 51 (2002).

\bibitem{lattimer}
J.M. Lattimer and M. Prakash, {\it ApJ}, {\bf 550}, 426 (2001).

\bibitem{BR}
O. Benhar and R. Rubino, {\it A\&A}, {\bf 434}, 247 (2005).

\bibitem{CF1991}
S. Chandrasekhar and V. Ferrari,  {\it Proc. R. Soc. Lond.}, {\bf A434}, 
449 (1991).

\bibitem{KS1992}
K.D. Kokkotas and B.F. Schutz, {\it MNRAS}, {\bf 255}, 119 (1992).

\bibitem{Paper2} 
S. Chandrasekhar and V. Ferrari, {\it Proc. R. Soc. London}, 
{\bf A432}, 247 (1991).

\bibitem{mnras}
O. Benhar, E. Berti and V. Ferrari, {\it MNRAS} {\bf 310}, 797 (1999).


\bibitem{EOSB}
V.R. Pandharipande, {\it Nucl. Phys.} {\bf A178}, 123 (1971).

\bibitem{EOSL}
V.R. Pandharipande and R.A. Smith, {\it Phys. Lett.} {\bf B59}, 15 (1975).


\bibitem{AK}
N. Andersson and K.D. Kokkotas, {\it MNRAS}, {\bf 299}, 1059 (1998).

\bibitem{astero}
O. Benhar, V. Ferrari and L. Gualtieri, {\it Phys. Rev. D}, {\bf 70}, 
124015 (2004)

\bibitem{AK2}
N. Andersson and K.D. Kokkotas, {\it MNRAS}, {\bf 320}, 307 (1999).

\bibitem{strange1}
A.R. Bodmer, {\it Phys. Rev. D}, {\bf 4}, 1601 (1971).

\bibitem{strange2}
E. Witten, {\it Phys.Rev. D}, {\bf 30}, 272 (1984).

\bibitem{Kyoto2002}
V.  Ferrari, G. Miniutti and J. A. Pons, {\it Class. Quant. Grav.}, {\bf 20}, 
8841 (2003).

\bibitem{DIS1}
T. Damour, B. R. Iyer and B. S. Sathyaprakash, {\it Phys. Rev. D} {\bf 57}, 885 (1998).



\end{thebibliography}
\end{document}